\documentclass[11pt]{article}
\usepackage{amsfonts,amsthm,amsmath,amssymb,latexsym,comment}
\usepackage{setspace}
\usepackage{verbatim}
\usepackage{amsmath}
\usepackage{amssymb}
\usepackage{color}
\usepackage{epsfig}
\usepackage{xspace}
\usepackage{graphics}
\usepackage{floatflt}
\usepackage{wrapfig}
\usepackage{url}

\newcommand{\remove}[1]{}

\newcommand{\be}{\begin{itemize}}
\newcommand{\ee}{\end{itemize}}
\newcommand{\bn}{\begin{enumerate}}
\newcommand{\en}{\end{enumerate}}

\newcommand{\bp}{\begin{proposition}}
\newcommand{\ep}{\end{proposition}}
\newcommand{\bl}{\begin{lemma}}
\newcommand{\el}{\end{lemma}}
\newcommand{\bco}{\begin{corollary}}
\newcommand{\eco}{\end{corollary}}
\newcommand{\bt}{\begin{theorem}}
\newcommand{\et}{\end{theorem}}
\newcommand{\bpr}{\begin{proof}}
\newcommand{\epr}{\end{proof}}
\newcommand{\bd}{\begin{definition}}
\newcommand{\ed}{\end{definition}}

\newcommand{\beqn}{\begin{centeqn}}
\newcommand{\eeqn}{\end{centeqn}}

\newcommand{\bleqn}[1]{\begin{centlabeqn}{#1}}
\newcommand{\eleqn}{\end{centlabeqn}}

\newcommand{\bc}{\begin{center}}
\newcommand{\ec}{\end{center}}

\newcommand{\ms}{\medskip}
\renewcommand{\ss}{\smallskip}

\newcommand{\bfg}{\begin{figure}}
\newcommand{\efg}{\end{figure}}

\newenvironment{centeqn}	
   {{\ss\\ \hspace*{\fill}}} 
   {\hspace*{\fill}\ss\\}

\newsavebox{\EqnLabel}
\newenvironment{centlabeqn}[1]		
   {\sbox{\EqnLabel}{#1}
    {\\ \hspace*{\fill}}
   } 
   {\hfill{\makebox[0in][r]{\usebox{\EqnLabel}}}\ms\\}

\newenvironment{centeqn-nbsp} 
   {{\ms\\ \hspace*{\fill}}}
   {\hspace*{\fill}}

\newcommand{\empi}{\textit}

\newcommand{\AND}{\bigwedge}

\newcommand{\OR}{\bigvee}

\newcommand{\oneton}{\{1 ... n\}}

\newcommand{\ar}{\rightarrow}	

\newcommand{\imp}{\Rightarrow}		
\newcommand{\ind}{\hspace*{3.0em}}
\newcommand{\ints}{\cap}

\newcommand{\la}[1]{\mbox{$\, \stackrel{#1}{\rightarrow} \,$}}

\newcommand{\pl}{\!\parallel\!}
\newcommand{\pj}{\ensuremath{\upharpoonright}}

\newcommand{\sat}{\models}

\newcommand{\sub}{\subseteq}
\newcommand{\un}{\cup}

\newenvironment{blst}{\begin{list}		
                       {--}
                       {\setlength{\topsep}{0em}
                        \setlength{\itemsep}{0em}
			\setlength{\leftmargin}{0.25in}
		     }}
                    {\end{list}}

\newcounter{levelone}
\newenvironment{nlst1}{\begin{list}	
                       {\arabic{levelone}.}
                       {\usecounter{levelone}
			\setlength{\topsep}{0em}
                        \setlength{\itemsep}{0em}
			\setlength{\leftmargin}{0.25in}
		     }}
                    {\end{list}}

\newcounter{leveltwo}

\newcommand{\Fa}{\ensuremath{\mathcal F}\xspace}

\newcommand{\IF}{\ensuremath{\mathbf{if}}\xspace}
\newcommand{\ELSE}{\ensuremath{\mathbf{else}}\xspace}
\newcommand{\THEN}{\ensuremath{\mathbf{then}}\xspace}

\newcommand{\RETURN}{\ensuremath{\mathbf{return}}\xspace}

\newcommand{\AP}{\ensuremath{\mathcal{AP}}\xspace}

\newcommand{\ACTLS}{\mbox{$\mathrm{ACTL}^*$}}  
\newcommand{\ATL}{\ensuremath{\mathrm{ATL}}\xspace}  
\newcommand{\CTL}{\ensuremath{\mathrm{CTL}}\xspace}  
\newcommand{\CTLS}{\ensuremath{\mathrm{CTL}^*}\xspace}

\newcommand{\A}{\mathrm{\mathsf{A}}}
\newcommand{\E}{\mathrm{\mathsf{E}}}

\newcommand{\U}{\mathrm{\mathsf{U}}}

\newcommand{\V}{\mathrm{\mathsf{V}}}

\newcommand{\X}{\mathrm{\mathsf{X}}}

\newcommand{\true}{\ensuremath{\mathsf{true}}\xspace}
\newcommand{\false}{\ensuremath{\mathsf{false}}\xspace}

\newcommand{\AG}{\textup{\textsf{AG}}}
\newcommand{\EG}{\textup{\textsf{EG}}}
\newcommand{\AF}{\textup{\textsf{AF}}}
\newcommand{\EF}{\textup{\textsf{EF}}}
\newcommand{\AU}{\textup{\textsf{AU}}}
\newcommand{\EU}{\textup{\textsf{EU}}}
\newcommand{\AV}{\mathrm{\mathsf{AV}}}
\newcommand{\EV}{\textup{\textsf{EV}}}

\newcommand{\AX}{\textup{\textsf{AX}}}
\newcommand{\EX}{\textup{\textsf{EX}}}

\remove{

\newtheorem{theorem}{Theorem}
\newtheorem{lemma}[theorem]{Lemma}
\newtheorem{proposition}[theorem]{Proposition}
\newtheorem{corollary}[theorem]{Corollary}

\newtheorem{definition}{Definition}

}

\newtheorem{theorem}{Theorem}

\newtheorem{lemma}{Lemma}
\newtheorem{proposition}[lemma]{Proposition}
\newtheorem{definition}{Definition}

\newtheorem{corollary}{Corollary}

\newcommand{\ie}{i.e.,\xspace}
\newcommand{\eg}{e.g.,\xspace}
\newcommand*{\kripkedef}{\ensuremath{(s_0, S, R, L)}\xspace}
\newcommand*{\kripkeprimedef}{\ensuremath{(s'_0, S', R',L')}\xspace}
\newcommand*{\FL}{\ensuremath{\mathit{sub}}}
\newcommand*{\cgsdef}{\ensuremath{(s_0, S, R, L,\sigma)}\xspace}
\newcommand*{\mucalc}{{\ensuremath{\mu}-calculus }}
\newcommand*{\Val}{\mathcal{V}}
\newcommand*{\ff}{\mathit{ff}}
\newcommand*{\ttt}{\mathit{tt}}
\newcommand*{\ATLA}{\ll\!A\!\gg}
\newcommand*{\ATLAneg}{\ll\!\Sigma - A\!\gg}

\newcommand*{\vphi}{\varphi}
\newcommand*{\repfor}{\ensuremath{\mathit{repair}}}
\newcommand*{\repmod}{\ensuremath{\mathit{model}}}
\newcommand*{\repair}{\textsc{Repair}\xspace}

\newcommand{\tSAT}{\ensuremath{\mathrm{3SAT}}\xspace}
\newcommand{\tcnf}{\ensuremath{\mathrm{3cnf}}\xspace}

\begin{document}


\title{Model and Program Repair via SAT Solving\thanks{This research was supported by NSF under Subcontract No. GA10551-124962}}
\author{
Paul Attie$^{\dag \S}$ and Jad Saklawi$^{\dag}$\\ \\
$^{\dag}$Department of Computer Science\\
American University of Beirut\\
$^{\S}$Center for Advanced Mathematical Sciences\\
American University of Beirut \\
\texttt{\{paul.attie, jad.saklawi\}@aub.edu.lb}
}

\maketitle

\begin{abstract}
We consider the following \emph{model repair problem}: given a
finite Kripke structure $M$ and a specification formula $\eta$ in
some modal or temporal logic, determine if $M$ contains a
substructure $M'$ (with the same initial state) that satisfies
$\eta$. Thus, $M$ can be ``repaired'' to satisfy the
specification $\eta$ by deleting some transitions.

We map an instance $(M, \eta)$ of model repair to a boolean formula
$\repfor(M,\eta)$ such that $(M, \eta)$ has a solution iff
$\repfor(M,\eta)$ is satisfiable. Furthermore, a satisfying assignment
determines which transitions must be removed from $M$ to generate a
model $M'$ of $\eta$. Thus, we can use any SAT solver to repair
Kripke structures.  Furthermore, using a complete SAT solver yields a
complete algorithm: it always finds a repair if one exists.

We extend our method to repair finite-state shared memory concurrent
programs, to solve the discrete event supervisory control problem
\cite{RW87,RW89}, to check for the existence of symmettric solutions
\cite{ES93}, and to accomodate any boolean constraint on the existence
of states and transitions in the repaired model.

Finally, we show that model repair is NP-complete for CTL, and logics
with polynomial model checking algorithms to which CTL can be reduced
in polynomial time.  A notable example of such a logic is
Alternating-Time Temporal Logic (ATL).
\end{abstract}


\section{Introduction and Motivation}

\label{sec:intro}
Counterexample generation in model checking produces an example
behavior that violates the formula being checked, and so facilitates
debugging the model. However, there could be many counterexamples, and
they may have to be dealt with by making different fixes manually, thus
increasing debugging effort.
In this paper we deal with all counterexamples
at once, by ``repairing'' the model: 
we present a method for automatically fixing Kripke
structures and shared memory concurrent programs with respect to CTL
\cite{EC82} and ATL \cite{AHK02} specifications. 

\textbf{Our contribution.}  
We first present a ``subtractive'' repair algorithm:
fix a Kripke structure only by removing transitions and states (roughly
speaking, those transitions and states that ``cause'' violation of the
specification).  If the initial state is not deleted, then the
resulting structure (or program) satisfies the specification.
We show that this algorithm is sound and relatively complete. 
An advantage of subtractive repair is that it does not introduce new
behaviors, and thus any missing 
(\ie not part of the formula being repaired against)
conjuncts of the specification that
are expressible in a universal temporal logic (no existential path
quantifier) are still satisfied (if they originally were).
Hence we can fix w.r.t. incomplete specifications.

We also extend the subtractive repair method in several directions: to
accommodate the addition of states and transitions, to solve the
discrete event supervisory control problem \cite{RW87,RW89}, to
accommodate arbitrary boolean constraints on the existence of states
and transitions in the repaired model, and 
to repair atomic
read/write shared memory concurrent programs.
Finally, we show that the model repair problem is NP-complete.


Formally, we consider the \emph{model repair problem}: given a Kripke
structure $M$ and a CTL or ATL formula $\eta$, does there exist a
substructure $M'$ of $M$ (obtained by removing transitions and states
from $M$) such that $M'$ satisfies $\eta$?  In this case, we say that
$M$ is \emph{repairable} w.r.t, $\eta$, or that a repair exists.

Our algorithm computes (in deterministic time polynomial in the size of $M$
times the size of $\eta$) a propositional formula $\repfor(M,\eta)$
such that $\repfor(M.\eta)$ is satisfiable iff $M$ contains
a substructure $M'$ that satisfies $\eta$. Furthermore, a satisfying
assignment for $\repfor(M.\eta)$ determines which transitions must be
removed from $M$ to produce $M'$. Thus, a single run of a complete SAT
solver is sufficient to find a repair, if one exists.
Our approach leverages the research investment in SAT solvers to 
attack the model repair problem.

Soundness of our repair algorithm means that the resulting $M'$ (if it
exists) satisfies $\eta$. Completeness means that if the initial
structure $M$ contains a substructure that satisfies $\eta$, then our
algorithm will find such a substructure, provided that a complete SAT
solver is used to check satisfaction of $\repfor(M.\eta)$.

While our method has a worst case running time exponential in the
number of global states, this occurs only if the underlying SAT solver
uses exponential time.  SAT-solvers have proved to be efficient in
practice, as demonstrated by the success of SAT-solver based tools
such as Alloy, NuSMV, and Isabelle/HOL.  The success of SAT solvers in
practice indicates that our method will be applicable to reasonable
size models, just as, for example, Alloy \cite{dnj02} is.

\textbf{Related work}.
The use of transition deletion to repair Kripke structures was
suggested in \cite{AE96,AE01} in the context of atomicity refinement: a
large grain concurrent program is refined naively (\eg by replacing a
test and set by the test, followed nonatomically by the set). In
general, this may introduce new computations (corresponding to ``bad
interleavings'') that violate the specification. These are removed by
deleting some transitions.

The use of model checking to generate counterexamples was suggested by 
Clarke et.~al.~\cite{CGMZ95} and Hojati et.~al.~\cite{HBK93}. 
\cite{CGMZ95} presents an algorithm for
generating counterexamples for symbolic model checking. \cite{HBK93}
presents BDD-based algorithms for generating counterexamples (``error
traces'') for both language containment and fair CTL model checking.
Game-based model checking \cite{SW91,SG03} provides a
method for extracting counterexamples from a model checking run. The
core idea is a \emph{coloring algorithm} that colors nodes in the
model-checking game graph that contribute to violation of the formula
being checked.

The idea of generating a propositional formula from a model checking
problem was presented in \cite{BCCZ99}. That paper considers LTL
specifications and bounded model checking: given an LTL formula $f$, a
propositional formula is generated that is satisfiable iff $f$ can be
verified within a fixed number $k$ of transitions along some path ($\E
f$). By setting $f$ to the negation of the required property,
counterexamples can be generated. Repair is not discussed.

Some authors \cite{JGB05,SJB05,Staber5b} have 
considered algorithms for solving the repair problem: given a program
(or circuit), and a specification, how to automatically modify the
program (or circuit), so that the specification is satisfied.
There appears to be no automatic repair method that is (1) complete (\ie
if a repair exists, then find a repair) for a full
temporal logic (\eg CTL, LTL), and (2) repairs all faults in a single
run, \ie deals
implicitly with all counterexamples ``at once.''
For example, Jobstmann et.~al.~\cite{JGB05} considers only one repair
at a time, and their method is complete only for invariants.  
In \cite{Staber5b}, the approach of \cite{JGB05} is 
extended so that multiple faults are considered at once, but at 
the price of exponential complexity
in the number of faults.

In~\cite{BEGL99} the repair problem for CTL is considered and solved
using adductive reasoning.
The method generates repair suggestions that must then be verified by
model checking, one at a time. In contrast, we fix all faults at once.

Antoniotti \cite{Ant96} has shown that the related problem of 
discrete event supervisory control is also NP-complete.


The rest of the paper is as follows.
Section~\ref{sec:prelim} provides brief technical preliminaries.
Section~\ref{sec:repair} is the core of the paper: it presents our
model repair method for CTL in detail, discuses how the method is modified
to handle ATL.
Section~\ref{sec:extensions} presents the various extensions discussed above.
Section~\ref{sec:app} presents several example applications of the method.
Section~\ref{sec:impl} discusses our implementation, including experimental performance data.
Section~\ref{sec:conc} discusses future work and concludes.
Appendix~\ref{app:example} presents a manual simplification of an example repair formula, 
Appendix~\ref{app:proofs} provides proofs for all theorems, and
Appendix~\ref{app:background} provides full technical preliminaries.




\section{Preliminaries}
\label{sec:prelim}

We assume basic of knowledge of \CTL\cite{Eme90,EC82} and
\ATL\cite{AHK02}.
The logic \CTL is given by the following grammar:
\[
\vphi ::= \true \mid \false \mid p \mid \neg \vphi \mid \vphi \land \vphi \mid \vphi \lor \vphi \mid 
          \AX \vphi \mid \EX \vphi \mid \A[\vphi \V \vphi] \mid \E[\vphi \V \vphi]
\]
where $p \in \AP$, a set of atomic propositions.
The semantics of a \CTL formula are given with respect to a Kripke
structure $M = \kripkedef$ where $s_{0}$ is the start state,$S$ is the
set of states, $R \sub S \times S$ is the transition relation and $L:S  \mapsto 2^{\AP}$
is the labeling function.
We use $M \sat \vphi$ to abbreviate $M,s_{0} \sat \vphi$.
We use the abbreviations 
$\A[\phi \U \psi]$ for $\neg \E[\neg \vphi \V \neg \psi]$,
$\E[\phi \U \psi]$ for $\neg \A[\neg \vphi \V \neg \psi]$,
$\AF \vphi$ for $\A[\true \U \vphi]$,
$\EF \vphi$ for $\E[\true \U \vphi]$,
$\AG \vphi$ for $\A[\false \V \vphi]$,
$\EG \vphi$ for $\E[\false \V \vphi]$.

\begin{definition}[Formula expansion]
\label{closure}
Given a CTL formula $\vphi$, its set of subformulae $sub(\vphi)$ is defined as follows:
\begin{itemize}
\item $sub(p) = {p}$ where $p$ is \true, \false, or an atomic proposition

\item $sub(\neg \vphi) = \{\neg \vphi\} \un sub(\vphi)$
\item $sub(\vphi  \land  \psi) = \{\vphi  \land  \psi\} \un sub(\vphi)
\un sub(\psi)$
\item $sub(\vphi  \lor \psi) = \{\vphi  \lor \psi\} \un sub(\vphi) \un
sub(\psi)$

\item $sub(\AX \vphi) = \{\AX \vphi\} \un sub(\vphi)$
\item $sub(\EX \vphi) = \{\EX \vphi\} \un sub(\vphi)$

\item $sub(\A[\vphi \V \psi]) = \{\A[\vphi \V \psi],\AX\A[\vphi \V \psi],\vphi  \lor
\AX\A[\vphi \V \psi] ,\psi  \land  (\vphi  \lor \AX\A[\vphi \V \psi])\}
\un sub(\vphi) \un sub(\psi)$

\item $sub(\E[\vphi \V \psi]) = \{\E[\vphi \V \psi],\EX\E[\vphi \V \psi],\vphi  \lor
\EX\E[\vphi \V \psi] ,\psi  \land  (\vphi  \lor \EX\E[\vphi \V \psi])\}
\un sub(\vphi) \un sub(\psi)$

\end{itemize}
\end{definition}

The logic \ATL is given by the following grammar:
\[
\vphi ::= \true \mid \false \mid p \mid \neg \vphi \mid \vphi \land \vphi \mid \vphi \lor \vphi \mid 
          \ATLA X \vphi \mid \ATLA[\vphi \V \vphi] 
\]
where $p \in \AP$, $A \subseteq \Sigma$. $\Sigma$ denotes the set of players.
$\ATLA \vphi$ holds iff the players in $A$ have a collective strategy
to enforce the truth of $\vphi$.

\section{The Model Repair Problem}
\label{sec:repair}


Given Kripke structure $M$ and a specification formula $\vphi$, we
consider the problem of removing parts of $M$, resulting in a
substructure $M'$ such that $M' \sat \vphi$. 


\begin{definition}[Substructure]
Given a Kripke structure $M = \kripkedef$ and a structure 
$M' = \kripkeprimedef$ we say that $M \sub M'$ iff
$S \sub S'$, $s_0 = s'_0$, $R' \sub R$, and $L' = L \pj S'$.
\end{definition}

\begin{definition}[Repairable]
\label{def:fixable}
\label{def:repairable}
Given Kripke structure $M = \kripkedef$ and a formula $\eta$.
$M$ is \emph{repairable} with respect to $\eta$ if there exists a 
Kripke structure $M' = \kripkeprimedef$ such that $M'$ is total, $M' \sub M$,
and $M',s_0 \sat \eta$.
\end{definition}
Recall that a Kripke strucutre is total iff every state has at least
one outgoing transition.

\begin{definition}[Model Repair Problem]
Given a Kripke structure $M = \kripkedef$, and a formula $\eta$,
the repair problem is to decide if $M$ is repairable with respect to $\eta$.
\end{definition}

The model repair problem is defined for any temporal or modal logic
for which the $\models$ relation is defined, e.g \mucalc, CTL*, CTL,
etc. So, for example, we speak of the model repair problem for CTL
(CTL model repair for short).  An instance of model repair is then the
pair $(M, \varphi)$.

\subsection{Complexity of the Model Repair Problem}
\label{sec:repair_problem}

\begin{theorem}
\label{thm:CTL-repair-complexity}
  The model repair problem for CTL is NP-complete.
\end{theorem}

\begin{corollary}
\label{cor:CTL-repair-complexity}
Let L be any temporal logic interpreted in Kripke structures such that
(1) model checking for L is in polynomial time, and
(2) there exists a polynomial time reduction from CTL model checking
  to L model checking.
Then the model repair problem for L is NP-complete.
\end{corollary}

An immediate consequence is that model repair for alternating-time
temporal logic (ATL) is NP-complete.

\subsection{CTL Model Repair using SAT solvers}
\label{sec:repair_sat}

Given an instance of model repair $(M,\eta)$, where $M = \kripkedef$ and $\eta$ is a
CTL formula, 
we define a propositional formula $\repfor(M,\eta)$ such that 
$\repfor(M,\eta)$ is satisfiable iff $(M,\eta)$ has a
solution. $\repfor(M,\eta)$ is defined over the following propositions:  
\begin{enumerate}
\item $E_{s,t}: (s,t) \in R$
\item $X_{s,\psi}: s \in S, \psi \in \FL(\eta)$
\item $X_{s,\psi}^n: s \in S, 0 \le n \le |S|$, and $\psi \in \FL(\eta)$ has the form
$\A[\vphi \V \vphi']$ or $\E[\vphi \V \vphi']$
\end{enumerate}
The meaning of $E_{s,t}$ is that the transition $(s,t)$ is retained in
the fixed model $M'$ iff  $E_{s,t}$ is assigned $\ttt$ (``true'') by the
satisfying valuation $\Val$  for $\repfor(M,\eta)$. 
The meaning of $X_{s,\psi}$ is that $\psi$ holds in state $s$.
$X_{s,\psi}^n$ is used to propagate release formula ($\AV$ or $\EV$)
for as long as necessary to determine their truth, \ie $|S|$ in the
worst case. 

A solution for satisfiability of $\repfor(M,\eta)$, \eg as given
by a SAT solver, gives directly a solution to model repair. Denote
this solution by \linebreak $\repmod(M,\Val)$. Then $\repmod(M,\Val) =
\kripkeprimedef$, where $R' = \{ (s,t) | \Val(E_{s,t}) = \ttt \}$,
$S'$ consists of all states reachable from $s_0$ via paths of
transitions in $R'$, and $L' = L \pj S'$.
Note that $\repmod(M,\Val)$ does not depend directly on $\eta$.

Essentially, $\repfor(M,\eta)$ encodes all of the usual local
constraints, \eg $\AX\vphi$ holds in $s$ iff $\vphi$ holds in
all successors of $s$. We modify these however, to take transition
deletion into account. So, the local constraint for $\AX$ becomes
$\AX\vphi$ holds in $s$ iff $\vphi$ holds in
all successors of $s$ \emph{after} transitions have been deleted (to
effect the repair). More precisely, instead of 
$X_{s,\AX\vphi} \equiv \AND_{t \mid s \ar t} X_{t,\vphi}$, 
we have 
$X_{s,\AX\vphi} \equiv \AND_{t \mid s \ar t} (E_{s,t} \imp X_{t,\vphi})$.
Here $s \ar t$ abbreviates $(s,t) \in R$.
The other modalities ($\EX, \AV, \EV$) are treated similarly.
We deal with $\AU, \EU$ by reducing them to $\EV, \AV$ using
duality. We require that the repaired structure $M'$ be total by requiring
that every state has at least one outgoing transition.

\newcommand{\vsp}{\vspace{1.0ex}}
\begin{definition}[$\repfor(M,\eta)$]
\label{def:repfor}
\noindent
Let $M = \kripkedef$ be a Kripke structure and $\eta$ a CTL formula.
Let $s \ar t$ abbreviate $(s,t) \in R$.
$\repfor(M,\eta)$ is the conjunction of all the propositional formulae listed
below. These are grouped into sections, where each section deals with one issue,
\eg propositional consistency. $s,t$ implicitly range over $S$.
Other ranges are explicitly given.


\vsp
$M'$ satisfies $\eta$: $X_{s_0,\eta}$

\vsp
\noindent
$M'$ is total, \ie each state has an outgoing
transition

\ind for all $s \in S:  \OR_{t \mid s \ar t} E_{s,t}$

\vsp
\noindent
Propositional labeling

for all $p \in \AP \ints L(s)$: $X_{s,p}$

for all $p \in \AP - L(s)$ : $\neg X_{s,p}$

\vsp
\noindent
Propositional consistency

for all $\neg \vphi \in \FL(\eta)$: $X_{s,\neg\vphi} \equiv \neg X_{s,\vphi}$

for all $\vphi\lor\psi \in \FL(\eta)$:  $X_{s,\vphi\lor\psi} \equiv X_{s,\vphi}
\lor X_{s,\psi}$

for all $\vphi\land\psi \in \FL(\eta)$: $X_{s,\vphi\land\psi} \equiv X_{s,\vphi} \land X_{s,\psi}$

\vsp
\noindent
Nexttime formulae

for all $\AX\vphi \in \FL(\eta)$: $X_{s,\AX\vphi} \equiv \AND_{t \mid s \ar t} (E_{s,t} \imp X_{t,\vphi})$

for all $\EX\vphi \in \FL(\eta)$: $X_{s,\EX\vphi} \equiv \OR_{t \mid s \ar t} (E_{s,t} \land X_{t,\vphi})$

\vsp
\noindent
Release formulae. Let $n = |S|$, \ie the number of states in $M$.

for all $\A[\vphi \V \psi] \in \FL(\eta)$: $X_{s,\A[\vphi \V \psi]} \equiv X^n_{s,\A[\vphi \V \psi]}$

for all $\A[\vphi \V \psi] \in \FL(\eta)$, $m \in \oneton$: \\
\ind $X_{s,\A[\vphi \V \psi]}^m \equiv 
     X_{s,\psi} \land 
     (X_{s,\vphi}
      \lor 
     \AND_{t \mid s \ar t} (E_{s,t} \imp X_{t,\A[\vphi \V \psi]}^{m-1}))$

for all $\A[\vphi \V \psi] \in \FL(\eta)$: $X_{s,\A[\vphi \V \psi]}^{0} \equiv X_{s,\psi}$ 

for all $\E[\vphi \V \psi] \in \FL(\eta)$: $X_{s,\E[\vphi \V \psi]} \equiv X^n_{s,\E[\vphi \V \psi]}$

for all $\E[\vphi \V \psi] \in \FL(\eta)$, $m \in \oneton$: \\
\ind $X_{s,\E[\vphi \V \psi]}^m \equiv
     X_{s,\psi} \land 
     (X_{s,\vphi}
      \lor
     \OR_{t \mid s \ar t} (E_{s,t} \land X_{t,\E[\vphi \V \psi]}^{m-1}))$

for all $\E[\vphi \V \psi] \in \FL(\eta)$: $X_{s,\E[\vphi \V \psi]}^0 \equiv X_{s,\psi}$ 


\end{definition}
We handle the ``$\vphi$ releases $\psi$'' modality $[\vphi \V \psi]$ as follows.
Along each path, either (1) a state is reached where $[\vphi \V \psi]$
is discharged ($\vphi \land \psi$), or (2) $[\vphi \V \psi]$ is shown to be false ($\neg \vphi
\land \neg \psi$), or (3) some state eventually repeats. In case (3), we
know that release also holds along this path.
Thus, by expanding the release modality up to $n$ times, where
$n$ is the number of states in the original structure $M$, we ensure
that the third case holds if the first two have not yet resolved the
truth of $(\vphi \V \psi)$ along the path in question.
To carry out the expansion correctly, we use a version of 
$X_{s,\A[\vphi \V \psi]}$ that is superscripted with an integer between
$0$ and $n$. This imposes a ``well foundedness'' on the 
$X^m_{s,\A[\vphi \V \psi]}$ propositions, and prevents for example, a
cycle along which $\psi$ holds in all states and yet the 
$X_{s,\A[\vphi \V \psi]}$ are assigned false in all states $s$ along
the cycle. 

Note that the above requires all states, even those 
rendered unreachable by transition deletion, to have some outgoing transition.
This ``extra'' requirement on the unreachable states does not affect
the method however, since there will actually remain a satisfying assignment
which allows unreachable state to retain all their outgoing
transitions, if some $M' \sub M$  exists that satisfies $\eta$.
For $s$ unreachable from $s_0$ in $M'$, assign the value to $X_{s,\vphi}$
that results from model checking $M',s \sat \vphi$. This gives a
consistent assignment that satsifies $\repfor(M,\eta)$. Clearly,  $X_{s,\vphi}$
does not affect $X_{s_0,\eta}$ since $s$ is unreachable from $s_0$.

In each state $s \in S$, there are $O(|\eta| \times |S|)$ formulae to
check, each of which has length $O(d)$, where $d$ is the maximum
number of succesors that any state in $S$ has.
The sum of lengths of all these formulae is 
$O(|\eta| \times |S|^2 \times d)$.
The propositional labelling formulae add $O(|S| \times |\AP|)$ length,
and so the size of $\repfor(M,\vphi)$ is 
$O(|\eta| \times |S|^2 \times d + |S| \times |\AP|)$,
and so is polynomial in the size of $(M,\eta)$. Clearly,
$\repfor(M,\eta)$ can be constructed in polynomial time.
Figure~\ref{fig:model-repair} presents our model repair algorithm,
$\repair(M,\vphi)$, which we show  is sound, and complete
provided that a complete SAT-solver is used.
Recall that we use $\repmod(M,\Val)$ to denote the structure $M'$ derived from
the repair of $M$ w.r.t. $\eta$, \ie $M' = \kripkeprimedef$, where 
$R' = \{(s,t) | \Val(E_{s,t}) = \ttt \}$, $S'$ consists of all states reachable from
$s_0$ via paths of transitions in $R'$, and $L' = L \pj S'$.

\begin{theorem}[Soundness]
\label{thm:sound}
Let $M = \kripkedef$ be a Kripke structure, $\eta$ a CTL formula,
and $n = |S|$. 
Suppose that $\repfor(M, \eta)$ is satisfiable and that 
$\Val$ is a satisfying truth assignment for it.
Let $M' = \repmod(M,\Val)$, 
Then for all reachable states $s \in S'$ and CTL formulae $\xi \in \FL(\eta)$:\\
\ind $\Val(X_{s,\xi}) = \ttt \text{ iff } M',s \sat \xi$ and\\
\ind for $m \in \oneton:$ $\Val(X^m_{s,\xi}) = \ttt \text{ iff } M',s \sat \xi$.
\end{theorem}

\begin{corollary}[Soundness] 
\label{cor:sound}
If $\repair(M,\eta)$ returns a structure $M' =\kripkeprimedef$, then 
(1) $M'$ is total, (2) $M' \sub M$, (3) $M',s_0 \sat \eta$, and 
(4) $M$ is repairable.
\end{corollary}

\begin{theorem}[Completness]
\label{thm:completeness}
If $M$ is repairable with respect to $\eta$ then $\repair(M,\eta)$ returns a 
Kripke structure $M''$ such that $M''$ is total, $M'' \sub M$, and $M'', s_0 \sat \eta$.
\end{theorem}

Since $M'$ results by removing transitions and unreachable states from
$M$, the relation mapping each state in $M'$ to ``itself'' in $M$ is a
simulation relation \cite{GL94} from $M'$ to $M$. 
Hence the following, where $\ACTLS$ \cite{GL94} is the universal fragment (no
existential path quantifier) of $\CTLS$, and clause (2) follows
from \cite{GL94}.
\begin{proposition}
\label{prop:simulation}
If $\repair(M,\eta)$ returns a structure $M'$, then (1) there is a
simulation relation from  $M'$ to $M$, and (2) for all \ACTLS\ formulae
$f$, $M \sat f$ implies $M' \sat f$.
\end{proposition}

\begin{figure}[th]
  \centering
  \framebox{
    \begin{minipage}{1.0\linewidth}
    \begin{tabbing}
    mmm\=mmm\=mmm\= \kill
       \>\underline{$\repair(M,\eta)$}:\\[1ex]
       \>model check $M, s_0 \sat \eta$;\\
       \>\IF successful, \THEN \RETURN $M$\\
       \>\ELSE\\
       \>   \>compute $\repfor(M, \eta)$ as given in Section~\ref{sec:repair};\\
       \>   \>submit $\repfor(M, \eta)$ to a sound and complete SAT-solver;\\
       \>   \>\IF the SAT-solver returns ``not satisfiable'' \THEN\\
       \>   \>   \> \RETURN ``failure''\\
       \>   \>\ELSE\\
       \>   \>   \>the solver returns a satisfying assignment $\Val$;\\
       \>   \>   \>\RETURN $M' = \repmod(M,\Val)$
    \end{tabbing}
    \end{minipage}
}

  \caption{The model repair algorithm.}
  \label{fig:model-repair}
\end{figure}

\subsection{ATL Model Repair using SAT solvers}
\label{sec:atl_repair_sat}

We adapt Definition~\ref{def:repfor} for ATL as follows.

We omit the conjuncts for $\AX \vphi$, $\EX \vphi$, $\A[\vphi \V
\psi]$, $\E[\vphi \V \psi]$, and add the following conjuncts. Here
$\sigma(s)$
is the player whose turn it is to move in state $s$.\\

\noindent
Nexttime formulae
\[
\text{ for all } \X\vphi \in \FL(\eta): X_{s, \ATLA\X\vphi} \equiv
\begin{cases}
\OR_{t \mid s \ar t} (E_{s,t} \land X_{t, \ATLA\vphi}) \text{ if } \sigma(s) \in A\\
 \AND_{t \mid s \ar t} (E_{s,t} \imp X_{t, \ATLA\vphi}) \text{ if } \sigma(s) \not\in A
\end{cases}
\]

\noindent
Release formulae. Let $n = |S|$ and $m \in \oneton$.
Then,  for all $\ATLA[\vphi \V \psi] \in \FL(\eta)$ we have the following conjuncts:

\noindent
$X_{s,\ATLA[\vphi \V \psi]} \equiv X^n_{s,\ATLA[\vphi \V \psi]}$

\noindent
$X_{s,\ATLA[\vphi \V \psi]}^m \equiv 
   X_{s, \ATLA\psi} \land 
   (X_{s, \ATLA\vphi}
    \lor 
   \OR_{t \mid s \ar t} (E_{s,t} \land X_{t,\ATLA[\vphi \V \psi]}^{m-1}))$
 if $\sigma(s) \in A$

\noindent
$X_{s,\ATLA[\vphi \V \psi]}^m \equiv
    X_{s, \ATLA\psi} \land
    (X_{s, \ATLA\vphi}
     \lor
    \AND_{t \mid s \ar t} (E_{s,t} \imp X_{t,\ATLA[\vphi \V \psi]}^{m-1}))$
if $\sigma(s) \not\in A$

\noindent
$X_{s,\ATLA[\vphi \V \psi]}^0 \equiv X_{s, \ATLA\psi}$

As in Definition~\ref{def:repfor}, the above formula encodes the
possibilities for valuation of $\eta$ and all its subformulae on $M$,
and the possible substructures resulting from deleting transitions
from $M$. We can still reduce until to release, since 
$\ATLA[\vphi \U \psi] \equiv$ $\ATLAneg[\neg \vphi \V \neg \psi]$
in turn-based synchronous games \cite{LMO07}. 


\section{Extensions of the Subtractive Repair Algorithm}
\label{sec:extensions}

We now present several extensions to the subtractive repair algorithm given in the previous section.

\subsection{Addition of States and Transitions}

The subtractive repair algorithm performs repair by deleting
transitions, with states being implicitly deleted if they become
unreachable. Let $M = (s_0, S, R, L)$ be a Kripke structure with
underlying set of atomic propositions $\AP$. By adding some states and
transitions to $M$ before performing repair, we can end up with a
substructure $M'$ that includes some of the added states. Thus, we
have \empi{addiditve repair}: repair performed by adding states and transitions,

Let $S^+$ be a finite set of states such that $S \ints S^+ =
\emptyset$, let $L^+$ be an extension of $L$ to $S \un S^+$, and let
$R^+$ be a subset of $(S \un S^+) \times (S \un S^+) - R$.  Let $M^+ =
(s_0, S \un S^+, R \un R^+, L^+)$. So, $S^+$ represents the states
that are added to $M$, and $R^+$ represents the transitions that are
added. Note that added transitions can involve only the original state
($S$), only the added state ($S^+$), or one state from each of $S$,
$S^+$. We now execute the algorithm subtractive repair algorithm of 
Figure~\ref{fig:model-repair}, \ie $\repair(M^+,\eta)$.

In practice, the added states and transitions would be determined
either manually, by the user of the repair tool, or mechanically using
heuristics. While it seems possible to modify the repair formula directly
to accommodate state/transition addition (\eg by introducing new
propositions for the added states and transitions), doing so does not
seem to be any better than adding to the structure $M$ and then
regenerating the repair formula using the existing Definition~\ref{def:repfor}.
Note that proposition~\ref{prop:simulation} no longer holds when we
add states and transitions.

\subsection{Discrete Event Supervisory Control}

In the well-know discrete event supervisory control problem (DESC)
\cite{RW87,RW89}, a Kripke structure is given in which the transitions
are labelled as ``controllable'' and ``not controllable''. The problem
is to delete (disable) only controllable transitions so that the resulting
structure satisfies a property, \eg expressed in CTL. We easily
subsume DESC when the property is
expressed in CTL as follows. Conjoin to the repair formula
$\repair(M,\eta)$ the transition propositions $E_{s,t}$ for all
uncontrollable transitions $s \ar t$. Thus, we submit the following
formula to the SAT solver: 
$\repair(M,\eta) \land (\AND_{(s \ar t) \text{is uncontrollable}} E_{s,t})$.
The resulting assignment
produced by the SAT solver must then assign $\ttt$ to all $E_{s,t}$
for all uncontrollable transitions $s \ar t$, and so none of these
transitions are deleted. By
Theorem~\ref{thm:completeness} (completeness), our repair method will
then find a solution that involves deleting only controllable
transitions, if such a solution exists. Thus, we subsume the discrete
event supervisory control problem.

\subsection{Generalized Boolean Constraints on Transition and State Deletion}

The reduction given above used only simple conjunctions of $E_{s,t}$
propositions.
We can conjoin arbitrary boolean formulae over the $E_{s,t}$ to 
$\repair(M,\eta)$, \eg 
$E_{s,t} \equiv E_{s',t'} \land E_{s',t'} \equiv E_{s",t"}$
adds the constraint that either all three transitions $s \ar t$, $s'
\ar t'$, $s'' \ar t''$ are deleted, or none are. This is useful in
enforcing atomic read/write semantics in shared memory, as discussed
below.

We can also add constraints on deleting states as follows. We can
introduce a proposition $N_s$ ($N$ for ``node'') for each each state $s$ with meaning
that $s$ is retained in the final model iff $N_s$ is assigned $\ttt$.
We now modify the clause for $M'$ being total to:
for all $s \in S:  N_s \implies \OR_{t \mid s \ar t} E_{s,t}$, and we
add as conjunct:
for all $s \in S: \neg N_s \implies (\AND_{t \mid s \ar t} \neg
E_{s,t}) \land (\AND_{t \mid t \ar s} \neg E_{t,s})$, 
that is, a nondeleted state must have some outgoing transition, and a
deleted state has no transitions, either incoming or outgoing.

Suppose we have a Kripke structure for two processes $P_1$ and $P_2$
executing some protocol, \eg mutual exclusion. We can both fix the
protocol and require the result to be symmetric in $P_1$ and $P_2$
(\ie the code for $P_2$ results from interchanging the process indices
1 and 2 in the code for $P_1$ \cite{AE98}) by adding the conjunct $N_s
\equiv N_t$ for every pair of symmetric state $s,t$, \ie such that $t$
results from $s$ by interchanging the process indices 1 and 2, and
likewise for symmetric transitions (start and end states are
symmetric).  Thus, we can check for the existence of symmetric
concurrent algorithms. Note that these more general constraints cannot
be dealt with by discrete event supervisory control, which only allows
to specify individual transitions as controllable or not, and does not
allow relating the deletion of one transition to the deletion of another.

\subsection{Concurrent Program Repair}

We now extend our approach to the repair of shared memory concurrent
programs $P = P_1 \pl \cdots \pl P_K$, where processes atomically read,  write
one shared variable at a time. We provide repair
w.r.t, CTL specifications.
We partition $\AP$ into $\AP_1,\ldots,\AP_K$, where 
$\AP_i$ consists of the atomic propositions that can only be written
by $P_i$ (but can be read by other processes). There are also shared
variables $x_1,\ldots,x_m$ (with finite domains) that can be read and written by all processes.


We use the atomic read/write notation introduced in \cite{AE96,AE01}
for atomic read/write programs.    Each process $P_i$ is a synchronization
skeleton \cite{EC82}, \ie a directed graph where the nodes are
\emph{local states} that determine a truth assignment for the
propositions in $\AP_i$, and the arcs between
nodes are labeled with guarded commands; the guard reads atomic
propositions of other processes and shared variables, the body is a
parallel assignment that updates shared variables. 

The atomic propositions in $\AP_i$
are consolidated into a single variable $L_i$ (the ``externally
visible location counter'') owned by $P_i$ (\ie written by $P_i$ and
read by other processes), so that the value of $L_i$ in $s_i$ is the
set of all propositions in $\AP_i$ that hold in $s_i$. 
$L_i$ provides incomplete information to other processes about the
current local state of $P_i$: when $P_i$ writes to $L_i$, its change
of local state is visible to other processes. When $P_i$ writes to a
shared variable $x$, or reads, then its change of local state is not
visible to other processes.
Since $L_i$ encodes location information, 
a single machine word is usually sufficient to
store $L_i$.

$(s_i, B \ar A, t_i)$ denotes an arc in $P_i$ from local state $s_i$ to local
state $t_i$ that is labeled with guarded command $B \ar A$.
The restrictions to atomic read/write syntax (cf.~Definition 3.1.4 in
\cite{AE01}) are that each arc $(s_i, B \ar A, t_i)$  of $P_i$ is
either:
\begin{blst}
\item \emph{unguarded and single-writing}: there is no guard (\ie $B$ is
  ``\true'') and $A$  either writes to $L_i$ (\ie $L_i$ has different
  values in $s_i$ and $t_i$, so its value in $t_i$ must be written
  into it by $A$) or it
  writes to a single shared variable $x$ (\ie has the form $x := c$,
  where $c$ is a value from the domain of $x$, in which case $L_i$ must have
  the same value in $s_i$ and $t_i$,).

\item \emph{single-reading and nonwriting}: there is no assignment (\ie $A$
  is ``skip''), $L_i$ has the same value in $s_i$ and $t_i$,
  and $B$ has the form $Q_j \in L_j$ where $Q_j \in
  \AP_j, \ j \ne i$ or the form $x = c$. We call such a form for
  $B$ a \emph{simple term}.
\end{blst}

A \emph{global state} $s$ is a tuple 
$\langle s_1,\ldots,s_K,v_1,\ldots,v_m \rangle$ where $s_i$ is the
current local state of $P_i$, and $v_j$ is the current value of shared
variable $x_j$. We write $s \pj i$ for the component of $s$ that gives
the local state of $P_i$.

An arc $arc = (s_i, B \ar A, t_i)$ of $P_i$ is \emph{enabled} in global state $s$
iff $s \pj i = s_i$ and $s(B) = \true$. Execution of 
$(s_i, B \ar A, t_i)$ in a global state $s$ where it is enabled
generates a transition $s \la{arc} t$, where $t$ results from $s$ by
changing the local state of $P_i$ from $s_i$ to $t_i$, and changing
the value of $x$ to $c$ if $A$ has the form $x := c$. 
In general, an arc can be enabled in several global states.
In the global state transition diagram $M$ generated by execution of $P$,
the set of all transitions generated by a single arc is called a
\emph{family}. 
We label every transition by the name of the family that it belongs to. 
Two different families do not intersect, since
their transitions have different labels, even if the transitions have
the same ``effect'' on the global state. This makes the technical
development more convenient and does not cause loss of generality.
Thus, the set of transitions in $M$ is partitioned into
families.

Let $P$ be a shared memory atomic read/write concurrent program, and
$\eta$ a CTL specification for $P$. We generate the global state
transition diagram $M = \kripkedef$ of $P$.
Suppose that $\repfor(M,\eta)$ has a satisfying assignment $\Val$, and
that $\Val(E_{s,t}) = \ff$ for some transition $(s,t)$ in $M$. Let
$\Fa$ be the family that $(s,t)$ belongs to, and $P_i$ be the process
in which the arc $arc$ generating $\Fa$ occurs.  To preclude executing
$arc$ in global state $s$, the repaired $P_i$ must detect that $s$ is actually the
current global state (and then not execute $arc$). This
requires that $P_i$ read enough externally visible location counters
$L_j, j \ne i$, and shared variables, so that it can determine a
pattern of assignment of values to these that is unique to $s$.  In
general, this may require that $P_i$ read several
location counters and shared variables atomically. 

We now have two cases, depending on $arc$.
First, suppose that $arc$ is unguarded and single-writing, Then we
cannot modify $arc$ to read any information without violating the
atomic read/write syntax restriction (effectively, $arc$ becomes a
test-and-set operation). We are thus left with two options: either
make $s$ unreachable, by deleting other transitions, or delete all the
transitions in $\Fa$. This can be expressed as 
$(\AND_{(s,t) \in \Fa} (\neg E_{s,t} \imp \neg r_s)) \lor (\AND_{(s,t) \in \Fa} \neg E_{s,t})$,
where $r_s$ is the ``reachability'' proposition given in Definition~\ref{def:repfor}.
The first disjunct states that deletion of $(s,t)$ requires that $s$
be unreachable. The second disjunct states that all transitions in
$\Fa$ are deleted. We add the above as a conjunct to $\repfor(M,\eta)$.

The second case is that $arc = (s_i, B \ar A, t_i)$ is single-reading
and nonwriting.  Since $B$ holds in $s$, the repair cannot allow $B$
to continue being used as the guard for $arc$, unless $s$ is made
unreachable (in which case $B$ is never evaluated in $s$), or the
entire family $\Fa$ is deleted, in which case the arc $arc$ is removed
from $P_i$. 
However, it is possible that a simple term other than $B$
could be used to effect a repair, namely one that holds in the initial states of all
transitions in $\Fa$ except for the transition $(s,t)$, \ie in all
states $s'$ such that $s' \ne s \land \exists t': (s',t') \in \Fa$,
and also does not hold in any other global state.  In
\cite{AE01}, an algorithm for finding a suitable simple term, if it
exists, is presented. Essentially, the algorithm checks all possible
simple terms (their number is $O(|M|)$).  While our approach is not
able to replace the guard $B$ by another guard, it is capable of
deleting unsuitable simple terms, by removing the family corresponding
to the arc in which the simple term is used as a guard.  This
encourages an experimental style, where we add extra arcs to the
synchronization skeletons in the initial program, if we think they may
contain suitable guards.  Since this does not increase the number of
local states of any process, nor the number of shared variables, the
number of global states is unaffected. Thus, we could even add arcs
for every possible simple term. Note however that 
Proposition~\ref{prop:simulation}
could be violated, as the additional arcs may induce
additional transitions in $M'$ that are not simulated by $M$.
The conclusion of the preceding discussion is that we use the same
idea for repair as we did for the unguarded and single-writing case,
namely add
$(\AND_{(s,t) \in \Fa} (\neg E_{s,t} \imp \neg r_s)) \lor (\AND_{(s,t) \in \Fa} \neg E_{s,t})$,
a conjunct to $\repfor(M,\eta)$, with the possibility that we add
``extra arcs'' before repairing, to increase the possibilities for the
repair.

\section{Examples}
\label{sec:app}

\subsection{Simple Example for CTL Model Repair}
\label{sec:sample_run}

\begin{figure}[t]
\begin{center}
\scalebox{0.39}{\input{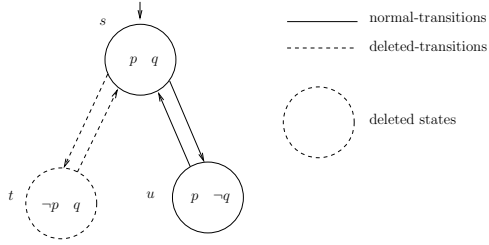}}
\end{center}
\caption{Input Kripke structure.}
\label{fig:input_struc}
\end{figure}

Consider the model
in 
Figure~\ref{fig:input_struc}
 and the formula $\eta = (\AG p
\lor \AG q) \land \EX p$.
Manual simplification of $\repfor(M,\eta)$ yields 
$\X_{s,\eta} \equiv \neg E_{s,t} \land E_{s,u}$,
so $\repair(M,\eta)$ will remove the
edge $(s,t)$ as shown.
Our implementation produces the following truth assignment:
\begin{verbatim}
A_A_A & s_u & ~s_t & u_s & t_s & ~u_5_0 & ~t_1_0 & s_10_0 & s_7_0 &
s_9_0 & s_0_0 & ~t_4_0 & ~s_5_0 & ~u_1_0 & ~s_6_0 & t_2_0 & ~u_2_0 &
t_0_0 & ~s_1_0 & ~t_3_0 & s_2_0 & s_8_0 & u_4_0 & ~s_3_0 & t_5_0 &
u_3_0 & s_4_0 & u_0_0
\end{verbatim}

The variable s\_t represents the edge from $s$ to $t$, etc.  Note that
s\_t is negated, indicating an assignment of $\ff$, \ie the edge should
be deleted, as required.

We also ran our implementation with repair formula $\AX p \land \AX
\neg p$. As expected, it returned ``unsatisfiable,'' indicating that
no repair exists.

\subsection{Barrier Synchronization Problem Repair}
\label{sec:barrier_sync}

\begin{figure}[t]
\begin{center}
\scalebox{0.6}{\input{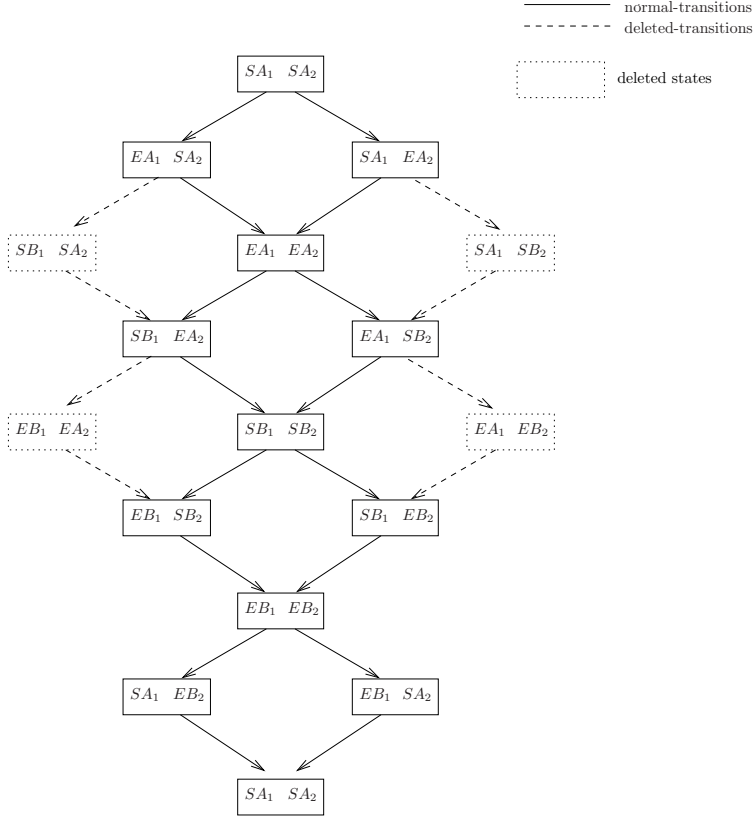}}
\end{center}
\caption{Barrier synchronization repair.}
\label{fig:barrier_sync}
\end{figure}

In this problem, each process $P_i$ is a cyclic sequence of two
terminating phases, phase $A$ and phase $B$. $P_ii$, $(i \in
\{1,2\})$, is in exactly one of four local states, $SA_{i}, EA_{i},
SB_{i}, EB_{i},$ corresponding to the start of phase $A$, then the end of
phase $A$, then the start of phase $B$, and then the end of phase $B$,
afterwards cycling back to $SA_{i}$.
 The CTL specification is the conjunction of the
following:
\begin{nlst1}
\item Initially both processes are at the start of phase $A$:
   $SA_{1}  \land  SA_{2}$
\item $P_{1}$ and $P_{2}$ are never simultaneously at the start of
 different phases:\\
\ind $\AG ( \neg (SA_{1}  \land  SB_{2}))  \land  \AG (\neg ( SA_{2}  \land  SB_{1} ))$
\item $P_{1}$ and $P_{2}$ are never simultaneously at the end of 
different phases:\\
\ind $\AG ( \neg (EA_{1}  \land  EB_{2}))  \land  \AG (\neg ( EA_{2}  \land  EB_{1} ))$
\end{nlst1}
$(2)$ and $(3)$ together specify the synchronization aspect of the problem: $P_{1}$ can
never get one whole phase ahead of $P_{2}$ and vice-versa. 

The structure in Figure \ref{fig:barrier_sync} is repaired
by removing edges and states that cause the
violation of the synchronization rules $(2)$ and $(3)$\footnote{Note that the bottom $[SA_1 SA_2]$ state is the same as the top $[SA_1
SA_2]$ state, and is repeated only for clarity of the figure.}
Our implementation produced exactly this repair.
The repair formula in CNF contained 236 propositions and 
401 clauses.

\section{Implementation of the Repair Method}
\label{sec:impl}

We implemented the method in Python. Our implementation takes a Kripke
structure $M$ and CTL formula $\eta$ as input, generates 
$\repfor(M,\eta)$
as given by Definition~\ref{def:repfor},
converts it
to CNF, and then invokes the SAT solver zChaff.
The implementation is available at
\url{http://www.cs.aub.edu.lb/pa07/pca/Eshmun.html}.

Table~\ref{table:repair_results} gives performance figures for our
implementation, running on a 
PC with Pentium 4 CPU at 3.00GHz, and 512MB RAM.
For $M$, we generated transitions graphs
randomly, specifying the number of nodes $N$ and the probability $P$
that there is a transition from some given node to some other given
node. We used a constant probability $P = 0.1$.
In $M$, we used $\AP = \{p,q\}$, and the propositional labels were
generated randomly for each state.

We show the number of propositions and clauses in the CNF form 
of $\repfor(M,\eta)$, and the total time our implementation takes to 
produce a satisfying assignment. This shows a typically expected
increase with the number of nodes in the graph.

For $N = 20, 30$ we used $\eta = \AX\A[p \V q] \land \EX q$.
For $N$ = 40 to 80, we used $\eta = \A[p \V q]$.

\def\sep{0.0ex}

\vspace{-5ex}

\begin{table}[ht]
\caption{Model Repair Results}
\centering
\begin{tabular}{c c c c}
\hline\hline 
$N$ & ~~Propositions~~ & ~~Clauses~~ & ~~Time~~ \\ [0.5ex]
\hline
30 & 309 & 3506 & 2.437s \\
40 & 449 & 3986 & 3.563s \\
50 & 608 & 13909 & 9.228s \\
60 & 781 & 47665 & 31.223s \\
70 & 993 & 106136 & 1m52.231s\\
80 & 1183 & 174107 & 3m17.140s \\
\hline
\end{tabular}
\label{table:repair_results}
\end{table}

\vspace{-5ex}

\section{Conclusions}
\label{sec:conc}

We presented a method for repairing Kripke structures and concurrent
programs so that they satisfy a CTL formula $\eta$, by deleting
transitions that ``cause'' violation of $\eta$. Our method is sound,
and is complete relative to our transition deletion strategy.  We
address the NP-completeness of our model repair problem by translating
it (in polynomial time) into a propositional formula, such that a
satisfying assignment determines a solution to model repair. Thus, we
can bring SAT solvers to bear, which leads us to believe that our
method will apply to nontrivial structures and programs, despite the
NP-completeness. Unlike other methods, ours both fixes all
counterexamples at once, and is complete for temporal properties,
specifically full CTL.  We extended our method in various directions,
to allow addition of states and transitions, to solve discrete event
supervisory control, and to repair shared memory concurrent
programs. We also provided experimental results from our
implementation.


Future work includes application of our implementation to larger
examples and case studies, and extension to hierarchical Kripke
structures. Our implementation is useful in model construction, where
it provides a check that the constructed structure contains a
model.


\remove{
: our method
does not penalize ``extra'' transitions that cause incorrect behavior,
since these are deleted. Thus, it encourages a style of concurrent
algorithm development that ``overestimates' the amount of transitions
needed: as long as a correct model is contained 'within'' the actual
model, we will find a correct model. 
}


\small
\bibliographystyle{plain}
\bibliography{abbrev,ref}
\normalsize

\clearpage
\appendix

\section{Manual Simplification of the Repair Formula of the First
  Example}
\label{app:example}


We show how $\repfor(M,\eta)$.
for our first example in Section~\ref{sec:app}
is simplified manually.
We omit the clauses dealing with reachability.

$X_{s,\eta} \equiv X_{s,(\AG p \lor \AG q) \land \EX p}$

$X_{s,(\AG p \lor \AG q) \land \EX p} \equiv X_{s,\AG p \lor \AG q} \land \X_{s,\EX p}$

$X_{s,\AG p \lor \AG q} \equiv X_{s,\AG p} \lor X_{s,\AG q}$

\noindent
We start by solving for $X_{s, \AG p}$.

$X_{s,\AG p} \equiv X_{s,\AG p}^{3}$

$X_{s,\AG p}^{3} \equiv X_{s,p} \land (E_{s,t} \imp X_{t,\AG p}^{2}) \land (E_{s,u} \imp X_{u,\AG p}^{2})$

$X_{t,\AG p}^{2} \equiv X_{t,p} \land (E_{t,s} \imp X_{s,\AG p}^{1})$

$X_{u,\AG p}^{2} \equiv X_{u,p} \land (E_{u,s} \imp X_{s,\AG p}^{1})$

$X_{s,\AG p}^{1} \equiv X_{s,p} \land (E_{s,t} \imp X_{t,\AG p}^{0}) \land (E_{s,u} \imp X_{u,\AG p}^{0})$

$X_{t,\AG p}^{0} \equiv X_{t,p} \equiv \ff$

$X_{u,\AG p}^{0} \equiv X_{u,p} \equiv \ttt$

\noindent
By replacing $X_{s,p}$ etc. by their truth values, we can simplify the
above as follows.
It is more intuitive to work ``bottom up''

$X_{s,\AG p}^{1} \equiv \neg E_{s,t}$

$X_{u,\AG p}^{2} \equiv (E_{u,s} \imp X_{s,\AG p}^{1})$

$X_{u,\AG p}^{2} \equiv E_{u,s} \imp \neg E_{s,t}$

$X_{s,\AG p}^{3} \equiv \neg E_{s,t} \land (E_{s,u} \imp X_{u,\AG p}^{2})$

$X_{s,\AG p}^{3} \equiv \neg E_{s,t} \land (E_{s,u} \imp (E_{u,s} \imp \neg E_{s,t})) \equiv \neg E_{s,t}$

$X_{s,\AG p} \equiv \neg E_{s,t}$

\noindent
Symmetrically, we have:

$X_{s,\AG q} \equiv \neg E_{s,u}$

\noindent
It remains to solve for $X_{s,\EX p}$.

$X_{s,\EX p} \equiv (E_{s,t} \land X_{t,p}) \lor (E_{s,u} \land X_{u,p})$

\noindent
By replacing $X_{t,p}$ and $X_{u,p}$ by their values we get:

$X_{s,\EX p} \equiv (E_{s,t} \land \ff) \lor (E_{s,u} \land \ttt) \equiv E_{s,u}$

\noindent
Therefore, we now can solve for $X_{s,\eta}$ producing:

$\X_{s,\eta} \equiv 
   (\neg E_{s,t} \lor \neg E_{s,u}) \land E_{s,u} \equiv
   \neg E_{s,t} \land E_{s,u}$

The above solution implies that $\repair(M,\eta)$ will remove the
edge $(s,t)$ and all the resulting unreachable states as shown in figure~\ref{fig:input_struc}.

Note that for $\eta = (\AG p \lor \AG q)$, we obtain 
$\X_{s,\eta} \equiv (\neg E_{s,t} \lor \neg E_{s,u})$, which admits two satisfying
valuations, \ie removing either $(s,t)$ or $(s,u)$ produces the needed
repair.

\section{Proofs}
\label{app:proofs}

\subsection{Proof of Theorem~\ref{thm:CTL-repair-complexity}.}

\begin{proof}
Let $(M,\eta)$ be an arbitrary instance of the CTL model repair problem.

\textit{NP-membership}: Given a candidate solution $M'$, the condition $M' \subseteq M$
is easily verified in polynomial time. $M',s_0 \sat \eta$ is verified in linear time
using the CTL model checking algorithm of \cite{CES86}.

\textit{NP-hardness}:
We reduce \tSAT to CTL model repair.

Given a Boolean formula 
 $f = \AND_{1 \leq i \leq n}(a_{i} \lor b_{i} \lor c_{i})$ in \tcnf,
where $a_{i}, b_{i}, c_{i}$ are literals over the a set
$x_{1},\ldots, x_{m}$ of propositions, i.e each of $a_{i}, b_{i}, c_{i}$ is $x_{j}$ or $\neg x_{j}$, 
for some $j \in 1 \ldots m$.
We reduce $f$ to $(M,\eta)$ where
$M = \kripkedef$, 
$S = \{s_0,s_1,\ldots,s_m,t_1,\ldots,t_m\}$, and 
$R = \{ (s_0,s_1), \ldots, (s_0,s_m), (s_1,t_1), \ldots, (s_m,t_m) \}$, \ie
transitions from $s_0$ to each of $s_1,\ldots,s_m$, and a transition
from each $s_i$ to $t_i$ for $i = 1,\ldots,m$.
The underlying set $\AP$ of atomic propositions is 
$\{p_{1},\ldots p_{m}, q_{1}, \ldots, q_{m}\}$. 
These propositions are distinct from the $x_{1},\ldots,x_{m}$ used in the \tcnf formula $f$.
$L$ is given by:
\begin{itemize}
\item $L(s_{0}) = \emptyset$
\item $L(s_{j}) = p_{j}$ where $1 \leq j \leq m$
\item $L(t_{j}) = q_{j}$ where $1 \leq j \leq m$
\end{itemize}
$\eta$ is given by:
\[
\eta = \AND_{1 \leq i \leq n}(\vphi^{1}_{i} \lor \varphi^{2}_{i} \lor \varphi^{3}_{i})
\]
where: 
\begin{itemize}
\item if $a_{i} = x_{j}$ then $\vphi^{1}_{i} = \AG(p_{j} \imp \EX q_{j})$
\item if $a_{i} = \neg x_{j}$ then $\vphi^{1}_{i} = \AG(p_{j} \imp \AX \neg q_{j})$
\item if $b_{i} = x_{j}$ then $\vphi^{2}_{i} = \AG(p_{j} \imp \EX q_{j})$
\item if $b_{i} = \neg x_{j}$ then $\vphi^{2}_{i} = \AG(p_{j} \imp \AX \neg q_{j})$
\item if $c_{i} = x_{j}$ then $\vphi^{3}_{i} = \AG(p_{j} \imp \EX q_{j})$
\item if $c_{i} = \neg x_{j}$ then $\vphi^{3}_{i} = \AG(p_{j} \imp \AX \neg q_{j})$
\end{itemize}
Thus, if $a^{1}_{i} = x_{i}$, then the transition from $s_{i}$ to
$t_{i}$ (which we write as $s_i \ar t_i$  must be retained in $M'$, and if 
$a_{i} = \neg x_{i}$, then the transition $s_{i} \ar t_{i}$ must not appear in $M'$.
It is obvious that the reduction can be computed in polynomial time.\\
It remains to show that:

$f$ is satisfiable iff $(M,\eta)$ can be fixed. 
The proof is by double implication. 

\textit{$f$ is satisfiable implies that $(M,\eta)$ can be fixed}:
Let $\Val : \{x_{1},\ldots,x_{m}\} \mapsto \{\ttt,\ff\}$ be a satisfying truth 
assignment for $f$. Define $R'$ as follows.
$R' = \{(s_{0}, s_{i}), (s_{i},s_{i}), (t_{i},t_{i}) \mid 1 \leq i \leq m\} \un
\{(s_{i},t_{i}) \mid \Val(x_{i}) = \ttt\}$, \ie the transition $s_{i} \ar t_{i}$ is present in $M'$
if $\Val(x_{i}) = \ttt$ and $s_{i} \ar t_{i}$ is deleted in $M^{'}$ if $\Val(x_{i}) = F$.
We show that $M',s_{0} \sat \eta$.
Since $\Val$ is satisfying assignment, we have 
$\AND_{1 \leq i \leq n}(\Val(a_{i}) \lor \Val(b_{i}) \lor \Val(c_{i}))$.
Without loss of generality, assume that $\Val(a_{i}) = \ttt$ 
(similar argument for $\Val(b_{i}) = \ttt$ and $\Val(c_{i}) = \ttt)$.
We have two cases.
Case 1 is $a_{i} = x_{j}$. Then $\Val(x_{j}) = \ttt$, so $(s_{j},t_{j}) \in R'$.
Also since $a_{i} = x_{j}$, $\vphi^{1}_{i} = \AG(p_{j} \imp \EX q_{j})$.
Since $(s_{j},t_{j}) \in R'$, $M',s_{0} \sat \vphi^{1}_{i}$.
Hence $M',s_{0} \sat \eta$.
Case 2 is $a_{i} = \neg x_{j}$. Then $\Val(x_{j}) = \ff$, so $(s_{j},t_{j}) \not\in R'$.
Also since $a_{i} = \neg x_{j}$, $\vphi^{1}_{i} = \AG(p_{j} \imp \AX \neg q_{j})$.
Since $(s_{j},t_{j}) \not\in R^{'}$, $M',s_{0} \sat \vphi^{1}_{i}$.
Hence $M',s_{0} \sat \eta$.

\textit{$f$ is satisfiable follows from $(M,\eta)$ can be fixed}:
Let $M' = \kripkeprimedef$ be such that
$M' \subseteq M$, $M',s_{0} \sat \eta$.
We define a truth assignment $\Val$ as follows: $\Val(x_{j}) = \ttt$ iff $(s_{j},t_{j}) \in R'$.
We show that $\Val(f) = \ttt$, \ie 
$\Val(a_{i}) \lor \Val(b_{i}) \lor \Val(c_{i})$ for all $i = 1 \ldots n$.
Since $M,s_{0} \sat \eta$ we have
$M,s_{0} \sat \vphi^{1}_{i} \lor \vphi^{2}_{i} \lor \vphi^{3}_{i}$ for all $i = 1 \ldots n$.
Without loss of generality, suppose that $M,s_{0} \sat \vphi^{1}_i$ 
(similar argument for $M,s_{0} \sat \vphi^{2}_i$ and $M,s_{0} \sat \vphi^{3}_i$).
We have two cases.
Case 1 is $a_{i} = x_{j}$. Then $\vphi^{1}_{i} = \AG(p_{j} \imp \EX q_{j})$.
Since $M',s_{0} \sat \vphi^{1}_{i}$, we must have $(s_{j},t_{j}) \in R'$. Hence 
$\Val(x_{j}) = \ttt$ by definition of $\Val$. Therefore $\Val(a_{i}) = \ttt$. Hence
$\Val(a_{i}) \lor \Val(b_{i}) \lor \Val(c_{i})$.
Case 2 is $a_{i} = \neg x_{j}$. Then $\vphi^{1}_{i} = \AG(p_{j} \imp \AX \neg q_{j})$.
Since $M',s_{0} \sat \neg \vphi^{1}_{i}$, we must have $(s_{j},t_{j}) \not\in R'$. Hence 
$\Val(x_{j}) = \ff$. Therefore $\Val(a_{i}) = \ttt$.  Hence
$\Val(a_{i}) \lor \Val(b_{i}) \lor \Val(c_{i})$.
\end{proof}

\subsection{Proof of Corollary~\ref{cor:CTL-repair-complexity}.}
\begin{proof}
\textit{NP-membership}: guess the substructure $M'$ of $M$ and then check $M
\sat \eta$ in polynomial time using a polynomial time model checking
algorithm.\\
\textit{NP-hardness}: use the reduction from \tSAT to CTL model checking given
in the proof of Theorem~\ref{thm:CTL-repair-complexity}, and then use the
assumed reduction to L model checking.
\end{proof}

\subsection{Proof of  Theorem~\ref{thm:sound}.} 

\begin{proof}
We proceed by induction on the structure of $\xi$.
We sometimes write 
$\Val(X_{s,\xi})$ instead of $\Val(X_{s,\xi}) = \ttt$ and
$\neg \Val(X_{s,\xi})$ instead of $\Val(X_{s,\xi}) = \ff$.\\

Case $\xi = \neg \vphi$:\\
$\Val(X_{s,\xi}) = \ttt$ iff\\
$\Val(X_{s,\neg\vphi}) = \ttt$ iff (by propositional consistency clause
of Definition~\ref{def:repfor})\\ 
$\Val(X_{s,\vphi}) = \ff$ iff (by the induction hypothesis)\\
$not(M',s \sat \vphi)$ iff\\
$M',s \sat \neg\vphi$ iff\\
$M',s \sat \xi$\\

Case $\xi = \vphi \lor \psi$:\\
$\Val(X_{s,\xi}) = \ttt$ iff\\
$\Val(X_{s,\vphi \lor \psi}) = \ttt$ iff (by propositional
 consistency clause of Definition~\ref{def:repfor})\\ 
$\Val(X_{s,\vphi}) = \ttt$ or $\Val(X_{s,\psi}) = \ttt$ iff (by the induction hypothesis)\\
$(M',s \sat \vphi)$ or $(M',s \sat \psi)$ iff
$M',s \sat \vphi \lor \psi$ iff 
$M',s \sat \xi$\\

Case $\xi = \vphi \land \psi$:\\
$\Val(X_{s,\xi}) = \ttt$ iff\\
$\Val(X_{s,\vphi \land \psi}) = \ttt$ iff (by propositional
 consistency clause of Definition~\ref{def:repfor})\\ 
$\Val(X_{s,\vphi}) = \ttt$ and $\Val(X_{s,\psi}) = \ttt$ iff (by the induction hypothesis)\\
$(M',s \sat \vphi)$ and $(M',s \sat \psi)$ iff
$M',s \sat \vphi \land \psi$ iff 
$M',s \sat \xi$\\

Case $\xi = \AX\vphi$:\\
$\Val(X_{s,\xi}) = \ttt$ iff\\
$\Val(X_{s,\AX\vphi}) = \ttt$ iff\\
$\AND_{t \mid s \ar t} \Val(E_{s,t} \imp X_{t,\vphi}) = \ttt$ iff\\
$\AND_{t \mid s \ar t} \Val(E_{s,t}) = \ttt \imp  \Val(X_{t,\vphi}) = \ttt$ iff
(since $s$ is reachable by assumption, $E_{s,t}$ implies that $t$ also reachable, and apply the induction hypothesis)\\
$\AND_{t \mid s \ar t} (s,t)\in R' \imp M',t \sat \vphi$ iff\\
$M',s \sat \AX\vphi$ iff\\
$M',s \sat \xi$\\

Case $\xi = \EX\vphi$:\\
$\Val(X_{s,\xi}) = \ttt$ iff\\
$\Val(X_{s,\EX\vphi}) = \ttt$ iff\\
$\OR_{t \mid s \ar t} \Val(E_{s,t} \land X_{t,\vphi}) = \ttt$ iff\\
$\OR_{t \mid s \ar t} \Val(E_{s,t}) = \ttt \land \Val(X_{t,\vphi}) = \ttt$ iff
(since $t$ is reachable from $s$ by assumption, and apply the induction hypothesis)\\
$\OR_{t \mid s \ar t} (s,t)\in R' \land M',t \sat \vphi$ iff\\
$M',s \sat \EX\vphi$ iff\\
$M',s \sat \xi$\\

Case $\xi = \A[\vphi \V \psi]$:
We do the proof for each direction separatly.

\vsp
Left to right, \ie $\Val(X_{s,\A[\vphi \V \psi]}) \text{ implies } M',s \sat \A[\vphi \V \psi]$:\\
$\Val(X_{s,\A[\vphi \V \psi]})$ iff\\
$\Val(X^n_{s,\A[\vphi \V \psi]})$ iff\\
$\Val(X_{s,\psi} \land 
      (X_{s,\vphi} \lor \AND_{t \mid s \ar t}(E_{s,t} \imp X^{n-1}_{t, \A[\vphi \V \psi]})))$ iff\\
(since $\Val$ is a valuation function, and so distributes over boolean connectives)\\
$\Val(X_{s,\psi}) \land 
(\Val(X_{s,\vphi_{1}}) \lor (\AND_{t\mid s \ar t} \Val(E_{s,t}) \imp \Val(X^{n-1}_{t, \A[\vphi_{1} \V \psi]})))$ iff (by the induction hypothesis)\\
$M',s \sat \psi \land 
(M',s \sat \vphi \lor \AND_{t \mid s \ar t}((s,t) \in R' \imp \Val(X^{n-1}_{t,\A[\vphi \V \psi]}))$.\\

We now have two cases
\begin{enumerate}
\item $M',s \sat \vphi$. In this case, $M',s \sat\A[\vphi \V \psi]$, and so $M',s \sat \xi$.
\item $\AND_{t \mid s \ar t} (s,t) \in R' \imp \Val(X^{n-1}_{t,\A[\vphi \V \psi]})$.
\end{enumerate}
For case 2, we proceed as follows.
Let $t$ be an arbitrary state such that $(s,t) \in R'$. Then 
$\Val(X^{n-1}_{t,\A[\vphi \V \psi]})$. If we show that 
$\Val(X^{n-1}_{t,\A[\vphi \V \psi]}) \text{ implies } M',s \sat \A[\vphi \V \psi]$ then we are done, 
by CTL semantics.
The argument is essentially a repetition of the above argument for 
$\Val(X_{s,\A[\vphi \V \psi]}) \text{ implies } M',s \sat \A[\vphi \V \psi]$.\\
Proceeding as above, we conclude $M',t \sat \psi$ and one of the same
two cases as above:
\begin{itemize}
\item $M',t \sat \vphi$
\item $\AND_{u \mid t \ar u} (t,u) \in R' \imp \Val(X^{n-2}_{u,\A[\vphi \V \psi]})$
\end{itemize}
However note that, in case 2, we are ``counting down.'' Since we count
down for $n = |S|$, then along every path starting from $s$, either
case (1) occurs, which ``terminates'' that path, as far as valuation
of $[\vphi \V \psi]$ is concerned, or 
we will repeat a state before (or when) the
counter reaches 0. Along such a path (from $s$ to the repeated state), $\psi$ holds
at all states, and so $[\vphi \V \psi]$ holds along this path. We
conclude that $[\vphi \V \psi]$ holds along all paths starting in $s$,
and so $M',s \sat \A[\vphi \V \psi]$.

\vsp
Right to left, \ie 
$\Val(X_{s, \A[\vphi \V \psi]}) \text{ follows from } M',s \sat \A[\vphi \V \psi]$:\\
Assume that $M',s \sat \A[\vphi \V \psi]$ holds.
Hence 
$M',s \sat \psi \land 
(M',s \sat \vphi \lor \AND_{t \mid t \ar s} ((s,t) \in R' \imp M',t \sat \A[\vphi \V \psi]))$.
By the induction hypothesis, 
$\Val(X_{s,\psi}) \land 
(\Val(X_{s,\vphi}) \lor \AND_{t \mid t \ar s} ((s,t) \in R' \imp M',t \sat \A[\vphi \V \psi]))$.
We now have two cases 
\begin{enumerate}
\item $\Val(X_{s,\vphi})$. 
Since we have $\Val(X_{s,\psi}) \land \Val(X_{s,\vphi})$ we conclude
$\Val(X_{s,\A[\vphi \V \psi]})$, and so we are done.
\item $\AND_{t \mid s \ar t} (s,t) \in R' \imp M',t \sat \A[\vphi \V \psi]$
\end{enumerate}
For case 2, we proceed as follows.
Let $t$ be an arbitrary state such that $(s,t) \in R'$. 
Then $M',t \sat \A[\vphi \V \psi]$.
If we show that 
$\Val(X^{n-1}_{t,\A[\vphi \V \psi]}) \text{ follows from } M',t \sat \A[\vphi \V \psi]$ then 
we can conclude $\Val(X_{s, \A[\vphi \V \psi]})$ by Definition~\ref{def:repfor}.
Proceeding as above, we conclude $\Val(X_{t,\psi})$ and one of the
same two cases as above:
\begin{itemize}
\item $\Val(X_{t,\vphi})$, so by Definition~\ref{def:repfor}, $\Val(X^{n-1}_{t,\A[\vphi \V \psi]})$ holds.
\item $\AND_{u \mid t \ar u} (t,u) \in R' \imp \Val(X^{n-2}_{u,\A[\vphi \V \psi]})$
\end{itemize}
As before, in case 2 we are ``counting down.'' Since we count
down for $n = |S|$, then along every path starting from $s$, either
case (1) occurs, which ``terminates'' that path, as far as
establishment of $\Val(X_{t,\vphi})$ is concerned, or 
we will repeat a state before (or when) the
counter reaches 0. Along such a path (from $s$ to the repeated state,
call it $v$), $\psi$ holds at all states. 
By Definition~\ref{def:repfor}, $X^0_{v,\A[\vphi \V \psi]} \equiv X_{v,\psi}$.
From $M',v \sat \psi$ and the induction hypothesis, $\Val(X_{v,\psi})$
holds. Hence $X^0_{v,\A[\vphi \V \psi]}$ holds.
Thus, along every path starting from $s$, we reach a state $w$ such
that $\Val(X^{m}_{w,\A[\vphi \V \psi]})$ holds for some $m \in \{0,\ldots,n\}$.
Hence by Definition~\ref{def:repfor}, $\Val(X_{s, \A[\vphi \V \psi]})$ holds.\\

Case $\xi = \E[\vphi \V \psi]$: this is argued in the same way as the
above case for $\xi = \A[\vphi \V \psi]$, except that we expand along
one path starting in $s$, rather than all paths. The differences with
the $\A\V$ case are straightforward, and we omit the details. 
\end{proof}

\subsection{Proof of  Corollary~\ref{cor:sound}.} 

\begin{proof}
Let $\Val$ be the truth assignment for $\repfor(M,\eta)$ that was
returned by the SAT-solver in the execution of $\repair(M,\eta)$. 
Since the SAT-solver is assumed sound, 
$\Val$ is actually a satisfying assignment for $\repfor(M,\eta)$.
For (1), let $u$ be an arbitrary reachable state in $M'$.
Consider a path from $s_0$ to $u$. By definition of $\repfor(M,\eta)$,
we have $\Val(E_{s,t}) = \ttt$ for every transition $(s,t)$ along this path.
Hence $\Val(\OR_{v \mid u \ar v} E_{u,v}) = \ttt$. Hence $u$ has some
outgoing transition in $M'$.  
(2) holds by construction of $M'$, which is derived from $M$ by
deleting transitions and (subsequently) unreachable states.
For (3), note that $X_{s_0,\eta}$ is a conjunct of $\repfor(M,\eta)$
by definition of $\repfor(M,\eta)$. Hence $\Val(X_{s_0,\eta}) =
\ttt$. Hence, by Theorem~\ref{thm:sound}, $M',s_0 \sat \eta$.
Finally, (4) follows from (1)--(3) and Definition~\ref{def:repairable}.
\end{proof}

\subsection{Proof of Theorem~\ref{thm:completeness}.}

\begin{proof}
Assume that $M$ is repairable with respect to $\eta$. 
By Definition~\ref{def:repairable}, there exists a total substructure $M'$
of $M$ such that $M', s_0 \sat \eta$. We define a satisfying valuation
$\Val$ for $\repfor(M,\eta)$ as follows.

Assign $\ttt$ to $E_{s,t}$ for every edge $(s,t) \in R'$ and $\ff$ to
every $E_{s,t}$ for every edge $(s,t) \not\in R'$.
Since $M'$ is total, the ``$M'$ is
total'' section is satisfied by this assignment.

Assign $\ttt$ to $X_{s_0,\eta}$. 
Consider an execution of the CTL model checking algorithm of
\cite{CES86} for checking $M', s_0 \sat \eta$. This algorithm will
assign a value to every formula $\vphi$ in $\FL(\eta)$ in every reachable
state $s$ of $M'$. Set $\Val(X_{s,\vphi}$ to this value.
By construction of the \cite{CES86} model checking algorithm,
these valuations will satisfy all of the
constraints given in the ``propositional labeling,'' ``propositional
consistency,'' ``nexttime formulae,'' and ``release formulae''
sections of Definition~\ref{def:repfor}.
Hence all conjuncts of $\repfor(M,\eta)$ are assigned $\ttt$ by
$\Val$. Hence $\Val(\repfor(M,\eta)) = \ttt$, and so $\repfor(M,\eta)$
is satisfiable.

Now the SAT-solver used is assumed to be complete, and so will return
some satisfying assignment for $\repfor(M,\eta)$ (not necessarily
$\Val$, since there may be more than one satisfying assignment).
Thus, $\repair(M,\eta)$ returns a structure $M'$, rather than
``failure.'' By corollary~\ref{cor:sound}, $M''$ is total, $M'' \sub M$,
and $M'', s_0 \sat \eta$.
\end{proof}

\section{Technical Background}
\label{app:background}

\subsection{Computation Tree Logic}

Let $\AP$ be a set of atomic propositions. including the constants \true and \false.
We use \true, \false as ``constant'' propositions whose
interpretation is always the truth values $\ttt$, $\ff$, respectively.


 The logic \CTL \cite{EC82} is given by the following grammar:
 \[
 \vphi ::= \true \mid \false \mid p \mid \neg \vphi \mid \vphi \land \vphi \mid \vphi \lor \vphi \mid 
           \AX \vphi \mid \EX \vphi \mid \A[\vphi \V \vphi] \mid \E[\vphi \V \vphi]
 \]
 where $p \in \AP$.

The semantics of formulae are defined with respect to a Kripke structure.
\begin{definition}
\label{def:kripke}
A Kripke structure is a tuple $M = \kripkedef$ 
where $S$ is a finite state of states, $s_{0} \in S$ is a single initial state,
$R \subseteq S \times S$ is a transition relation,
and $L : S  \mapsto 2^{\AP}$ is a labeling function 
that associates each state $s \in S$ with a subset of atomic
propositions, namely those that hold in the state.  
\end{definition}

We assume that a Kripke structure $M = \kripkedef$ 
is total, \ie $\forall s \in S, \exists s' \in S: (s,s') \in R$.
A path in $M$ is a (finite or infinite) sequence of states, $\pi =s_{0}, s_{1}, \ldots$ 
such that $\forall i \geq 0: (s_i,s_{i+1}) \in R$.
A fullpath is an infinite path.

\begin{definition}
\label{def:kripke_semantics}
$M,s \sat \vphi$ means that formula $\vphi$
is true  in state $s$ of structure $M$ and
$M,s \not\sat \vphi$ means that formula $\vphi$
is false in state $s$ of structure $M$.
We define $\sat$ inductively as usual:
\begin{itemize}
\item $M,s \sat \true$
\item $M,s \not\sat \false$
\item $M,s \sat p  \text{ iff } p \in L(s)$ where atomic proposition $p \in \AP$

\item $M,s \sat \neg \vphi$  \text{ iff } $M,s \not\sat \vphi$
\item $M,s \sat \vphi \land \psi
           \text{ iff } M,s \sat \vphi \text{ and } M,s \sat \psi$
\item $M,s \sat \vphi \lor \psi
           \text{ iff } M,s \sat \vphi \text{ or } M,s \sat \psi$

\item $M,s \sat \AX \vphi \text{ iff }
            \text{ for all $t$ such that } (s,t) \in R: (M,t) \sat \vphi$
\item $M,s \sat \EX \vphi \text{ iff } 
            \text{ there exists $t$ such that } (s,t) \in R \text{ and }  (M,t) \sat \vphi$

\item $M,s \sat \A[\vphi \V \psi] \text{ iff }
           \text{ for all fullpaths $\pi = s_0,s_1,\ldots$  starting from $s=s_0$: }$\\
$\forall k \geq 0: 
   (\forall j < k: (M,s_j \not\sat \vphi ) \text{ implies } M,s_k \sat \psi$

\item $M,s \sat \E[\vphi \V \psi] \text{ iff } 
           \text{ for some fullpath $\pi = s_0,s_1,\ldots$  starting from $s=s_0$: }$\\
$\forall k \geq 0: 
   (\forall j < k: (M,s_j \not\sat \vphi) \text{ implies } M,s_k \sat \psi$
\end{itemize}
\end{definition}
We use $M \sat \vphi$ to abbreviate $M,s_{0} \sat \vphi$.
We introduce the abbreviations 
$\A[\phi \U \psi]$ for $\neg \E[\neg \vphi \V \neg \psi]$,
$\E[\phi \U \psi]$ for $\neg \A[\neg \vphi \V \neg \psi]$,
$\AF \vphi$ for $\A[\true \U \vphi]$,
$\EF \vphi$ for $\E[\true \U \vphi]$,
$\AG \vphi$ for $\A[\false \V \vphi]$,
$\EG \vphi$ for $\E[\false \V \vphi]$.







\subsection{Alternating-Time Temporal Logic}
\label{sec:ATL}

We review \emph{Alternating-Time Temporal Logic} (ATL)\cite{AHK02}.
ATL extends the existential and universal quantification over paths of
CTL by offering selective path quantification by a set of \emph{players}, \ie
paths along which the set of players can ``enforce'' the satisfaction
of a formula.  In general, ATL is interpreted over \emph{concurrent game
structures} where every state transition results from each player
choosing it's move, and then all players moving ``at the same time.'' 
There are also several kinds of restricted structures in which ATL can
be interpreted.
\emph{Turn-based synchronous} games are games where in each
step only one player makes a move, and the current player is
determined by the current state.  \emph{Moore synchronous} games are
games where the state space is partitioned according to the players,
and in each step, every player updates its own components of the state
independently of other players.  \emph{Turn-based asynchronous} are
games in which in each step only one player has a choice of moves and
that player is determined by a fair scheduler.  In this paper we
restrict ourselves to turn-based synchronous games.  The results
obtained still apply to other types of games since they can be reduced
to turn-based synchronous games in polynomial time \cite{AHK02}.


Let $\AP$ be a set of atomic propositions including the constants \true and \false.
Let $\Sigma$ denote the set of players.
 The logic \ATL is given by the following grammar:
 \[
 \vphi ::= \true \mid \false \mid p \mid \neg \vphi \mid \vphi \land \vphi \mid \vphi \lor \vphi \mid 
           \ATLA X \vphi \mid \ATLA[\vphi \V \vphi] 
 \]
 where $p \in \AP$, $A \subseteq \Sigma$

We use $M \sat \vphi$ to abbreviate $M,s_{0} \sat \vphi$.
We introduce the abbreviations 
$\ATLA[\phi \U \psi]$ for $\neg \ATLAneg [\neg \vphi \V \neg \psi]$, 
$\ATLA F \vphi$ for $\ATLA [\true \U \vphi]$,
$\ATLA G \vphi$ for $\ATLA [\false \V \vphi]$.

\remove{
\begin{definition}[ATL formula expansion]
  \label{def:atl_exp}
  Given an ATL path formula $\vphi$ of the form $\ATLA(\vphi \U \psi)$ or $\ATLA(\vphi \V \psi)$ , its expansion $exp(\vphi)$ is defined as follows:
\item $exp(\ATLA(\vphi \U \psi)) := \{\ATLA\vphi \U \psi,\ATLA\X(\vphi \U \psi),\vphi  \land  \ATLA\X(\vphi \U \psi) ,\psi  \lor (\vphi  \land  \ATLA\X(\vphi \U \psi))\}$
\item $exp(\ATLA(\vphi \V \psi)) := \{\ATLA(\vphi \V \psi),\ATLA\X\ATLA(\vphi \V \psi),\vphi  \lor \ATLA\X\ATLA(\vphi \V \psi),\psi  \land  (\vphi  \lor \ATLA\X\ATLA(\vphi \V \psi))\}$
\end{definition}
}

\begin{definition}[ATL formula subformulae]
\label{def:atl_sub}
Given an ATL formula $\vphi$, its subformulae $sub(\vphi)$ is defined as follows:
\begin{itemize}
\item $sub(p) := {p}$ where $p$ is true, false, or an atomic proposition
\item $sub(\vphi  \land  \psi) := \{\vphi  \land  \psi\} \un sub(\vphi) \un sub(\psi)$
\item $sub(\vphi  \lor \psi) := \{\vphi  \lor \psi\} \un sub(\vphi) \un sub(\psi)$
\item $sub(\ATLA\X \vphi) := \{\ATLA\X \vphi\} \un sub(\vphi)$
\item $sub(\ATLA(\vphi \V \psi)) := exp(\ATLA(\vphi \V \psi))
\un sub(\vphi) \un sub(\psi)$
\end{itemize}
\end{definition}

In \cite{AHK02}, the semantics of ATL is defined with respect to concurrent
game structures. Since we consider only 
turn-based synchronous game structures, we provide a 
(simpler) definition for ATL semantics with respect to 
turn-based synchronous game structures. 

\begin{definition}
\label{def:cgs}
A turn-based synchronous game structure is a tuple $M =$\linebreak $\cgsdef$
where $S$ is a finite state of states, $s_0$ is the single initial state,
$R \subseteq S \times S$ is a transition relation 
and $L : S  \ar 2^{\AP}$ is a labeling function 
that associates each state $s \in S$ with a subset of atomic
propositions, namely those that hold in the state. 
$\sigma$ is the turn function $\sigma: S  \mapsto \Sigma$ that
maps each state to a player (whose turn it is to make a move).
\end{definition}

We assume that each structure $M = \cgsdef$
is total, \ie $\forall s \in S, \exists s' \in S: (s,s') \in R$.
A path in $M$ is a (finite or infinite) sequence of states, $\pi =s_{0}, s_{1}, \ldots$ 
such that $\forall i \geq 0: (s_i,s_{i+1}) \in R$.
A fullpath is an infinite path.

For a path $\pi$ and a position $i \geq 0$, we use $\pi[i]$ to denote the $i$th
state of $\pi$.
A strategy for a player $a \in \Sigma$ is a mapping $f_{a}: S^{*} \mapsto S$ that assigns to every
finite path $\pi$ a successor state $s \in S$.
Given a state $s \in S$ and a set $A \subseteq \Sigma$ of players, an $A$-strategy 
$F_{A} = \{f_{a} \mid a \in A\}$ is a set of strategies, one for each player in $A$.
We define the outcomes of $F_{A}$ from $s$ to be the set $out(s,F_{A})$ of all fullpaths
that the players in $A$ can enforce when they follow the strategies in $F_{A}$, \ie
a fullpath $\pi = s_{0},s_{1}, \ldots$ is in $out(s,F_{A})$ if $s_{0} = s$ and for all
$i \geq 0$, if $a = \sigma(\pi[i])$ then 
$s_{i+1} = f_{a}(\pi[0,i])$.

\begin{definition}[ATL semantics]
\label{def:atl_semantics}
$M,s \sat \vphi$ means that 
$\vphi$ is true  in state $s$ of game structure $M = \cgsdef$.
$M,s \not\sat \vphi$ means that formula $\vphi$
is false in state $s$ of game structure $M$.
We define $\sat$ inductively as usual:
\begin{itemize}
\item $M,s \sat \true$
\item $M,s \not\sat \false$
\item $M,s \sat p \text{ iff } p \in L(s)$ where atomic proposition $p \in \AP$

\item $M,s \sat \neg \vphi$  \text{ iff } $M,s \not\sat \vphi$
\item $M,s \sat \vphi \land \psi
           \text{ iff } M,s \sat \vphi \text{ and } M,s \sat \psi$
\item $M,s \sat \vphi \lor \psi
           \text{ iff } M,s \sat \vphi \text{ or } M,s \sat \psi$

\item $M,s \sat \ATLA \X \vphi \text{ iff }$ there exists a set $F_{A}$ of strategies, 
one for each player in $A$, such that for all fullpaths $\pi \in out(s,F_{A})$, we have $M, \pi[1] \sat \vphi$

\item $M,s \sat \ATLA[\vphi \V \psi]$ iff there exists a set $F_{A}$ of strategies, 
one for each player in $A$, such that for all fullpaths $\pi \in out(s,F_{A})$:\\
$\forall k \geq 0: 
   (\forall j < k: (M,\pi[j] \not\sat \vphi ) \text{ implies } M,\pi[k] \sat \psi$

\end{itemize}
\end{definition}

\end{document}